\documentclass[doublecolumn,showpacs,preprintnumbers,amsmath,amssymb]{revtex4}
\usepackage{graphicx}
\usepackage{dcolumn}
\usepackage{bm}

\begin{document}
\title{Generalized cell/dual-cell transformation for percolation,
and new exact thresholds}

\author {Robert M. Ziff}
\affiliation{
{Michigan Center for
Theoretical Physics and Department of Chemical
Engineering, University of Michigan, Ann Arbor, MI USA 48109-2136} 
}


\date{\today}

\begin{abstract}
Suggested by  Scullard's recent star-triangle
relation for correlated bond systems, we propose a general
 ``cell/dual-cell" transformation, which
 allows in principle an infinite variety of lattices with exact 
 percolation thresholds
 to be generated.  We directly verify 
 Scullard's new  site percolation thresholds, and derive
the bond  thresholds for  
his ``martini" lattice ($p_c = 1/\sqrt{2}$)
and the ``A" lattice ($p_c =  0.625457\ldots$,  solution to
$p^5 - 4 p^4 + 3 p^3 + 2 p^2 - 1  = 0$).
We also present a precise Monte-Carlo test of the site
threshold for the ``A" lattice.
\end{abstract}
\maketitle
\bigskip

\vspace{0.5cm}
\section{Introduction}
Percolation refers to the process 
of the formation of long-range 
connectivity in random systems \cite{StaufferAharony}. 
It has wide-ranging applications to problems in physics
and engineering,
including conductivity and magnetism in random systems,
fluid flow in porous media, epidemics and clusters
in complex networks, and gelation in polymer systems. 
To study this phenomenon, one typically 
models the network by a regular lattice
made random
by independently making sites or bonds occupied
with a probability $p$.  At a critical threshold
$p_c$, for a given lattice and percolation type
(site, bond), percolation takes place.  Finding that
threshold exactly or numerically to high precision 
is essential to studying the percolation problem
on a particular lattice.

There are in fact relatively few lattices where $p_c$ is known
exactly, all in two dimensions, where a powerful
duality-crossing argument can be used.  These lattices include bond percolation on the
square lattice and site percolation on the triangular
lattice (both with $p_c = 1/2$), and a family of related lattices (bond-triangular,
bond-honeycomb, site-kagom\'e, site-(3,12$^2$)) whose non-trivial thresholds
can be found though the star-triangle transformation 
introduced by
Sykes and Essam \cite{SykesEssam}.  This transformation was successfully
applied to one other lattice (the ``bowtie" lattice,
bond percolation) by Wierman \cite{Wierman}.

Very recently Scullard has shown that the star-triangle
relation can be generalized to situations where the
bonds in the triangular unit have correlations
among them \cite{Scullard}.  He argued that a specific correlated system
could be represented by uncorrelated site percolation
on certain lattices, leading to exact percolation
thresholds for three new lattices: his so-called ``martini"
lattice, 
and two related lattices, which we call  
``A"  and ``B".  Scullard's proposed
thresholds are $p_c(\hbox{site}) = 0.764826...$ (the solution to 
$p^4-3p^3+1 = 0$),  $1/\sqrt{2}$, and $(\sqrt{5}-1)/2$,
respectively.
These rather startling results expand in a significant
way the number and types of lattices where exact
thresholds can be found.

In this paper, we show that Scullard's arguments
can be viewed in a more general way,
in terms of the net correlation between
the three vertices of a basic triangular cell.
 This correlation can be
created by an arbitrary network of independent
 or correlated bonds, including
groups of correlated bonds that effectively
represent sites in the system.  In this way, we derive
the bond thresholds for the lattices Scullard
considered (he already realized that the self-dual
``B" lattice has a threshold of 1/2), give an alternative
derivation of his site thresholds, and 
provide a prescription
to generate additional lattices where the threshold
can be found exactly.   Finally, we report on numerical simulations
to test Scullard's prediction of $p_c(\hbox{site})$ for the ``A" lattice.

\section{General criterion for percolation}

We consider a lattice that can be decomposed into 
a regular triangular array of identical triangular cells as shown in Fig.\ 1a, 
where the shaded triangles represent any network of bonds,
perhaps with correlations, that connect the three endpoints.
Randomly occupied sites can also be included
within the cell by incorporating triangles of 
correlated bonds.
That that array is triangular as in Fig.\ 1a
is not absolutely necessary (it need only be self-dual),
but we will consider only the triangular array in this paper.
The basic triangular cell has vertices $A$, $B$, 
and $C$, and we define $P(A,B,C) = $ the probability
that $A$, $B$ and $C$ are all connected, $P(A, B, \overline C)$
= the probablity that only $A$ and $B$ are connected, 
$P(\overline A, \overline B, \overline C) = $ the probability
that none of the three are connected, etc. 

We find that the general criterion for criticality is
simply
\begin{equation}
P(A,B,C) = P(\overline A, \overline B, \overline C)
\label{ziff}
\end{equation}
for an arbitrary triangular cell.
This formula is analogous to one that is
obtained  for
the triangular and honeycomb lattices
via the star-triangle transformation, but generalized
to apply to triangular cells with any number 
of bonds (including correlations), arranged
in a self-dual pattern such as that in Fig.\ 1a.

\begin{figure}
 \includegraphics[width=3in]{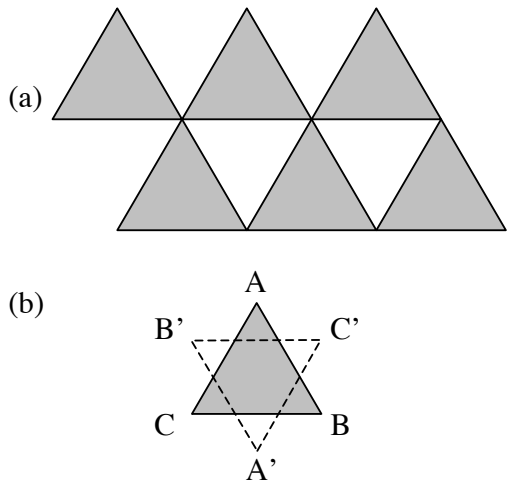} \caption{\label{fig1}
 (a) Decomposition of lattice into cells (shaded),
 (b) the dual-cell transformation.} \end{figure}

We derive this result using a generalized
``cell/dual-cell" transformation.
The development parallels that for the regular star-triangle
transformation, but kept in a more general context.
The dual system to Fig.\ 1a is also a regular triangular
array of non-overlapping triangles (rotated by 180$^\circ$).
Call the
vertices $A'$, $B'$, and $C'$ on the dual triangle,
as shown in Fig.\ 1b, and
let $P'(A',B',C')$ etc.\ represent the probabilities of 
connecting vertices on the dual triangular cells.  

For the two systems to be identical, rotate
the dual system so that $A = A'$, $B = B'$ and
$C = C'$.  Evidently, the two systems will have the same
connectivity between vertices if we have simultaneously
\begin{eqnarray}
P(A,B,C) &=& P'(A,B,C) \label{all} \\
P(A, B, \overline C) &=& P'(A, B, \overline C) \quad (\times 3)
\label{one} \\
P(\overline A, \overline B, \overline C) &=& P'(\overline A, 
\overline B, \overline C)  \label{none}
\end{eqnarray}
Note that for the two systems to have the same connectivity,
the underlying
bond occupancy of the dual cell will be different than that of the
original cell (in fact, $1 - p_i$ for a system of independent bonds),
but we don't need to specify that in detail here.

First we note that the two-point connectivity
relations (\ref{one}) are 
automatically satisfied as a result of duality.
We see
from Fig.\ 1b that if $A$ and $B$ connect on the 
original triangular cell, then $A'$ and $B'$ must also 
connect on the dual cell.  Likewise, duality also 
implies that $P(A,B,C) = P'(\overline A, 
\overline B, \overline C)$, and 
 $P(\overline A, \overline B, \overline C) = P'(A,B,C)$.  Then, from
Eqs.\ (\ref{all}) and (\ref{none}), the general condition
(1) follows. This indeed represents the critical
point because the two systems
have 
identical connectivities, and one is the dual
of the other.

Eq.\ (\ref{ziff}) is equivalent to 
the relation $P(v,h) + P(v,l,\overline h) + P(h, l, \overline v)
- P(\overline v, \overline l, \overline h) = 0$ given by
Scullard for a system of three correlated bonds,
where $v$, $h$, and $l$ are the vertical, horizontal,
and diagonal edges of the basic triangle drawn as a right
triangle.  From the identity 
$P(v,h) = P(v, h, l) + P(v, h, \overline l)$,
and associating $P(A,B,C) = P(v,h,l) + P(v, h, \overline l) +P(v,l,\overline h) + P(h, l, \overline v)$
and 
$P(\overline A, \overline B, \overline B) = P(\overline v, \overline l, \overline h)$,
 it follows that (\ref{ziff}) is satisfied.  Note that
 in the present work we consider systems with correlations
 more general than those produced by three correlated
 bonds.

\begin{figure}
 \includegraphics[width=3in]{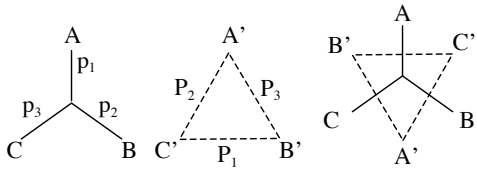} \caption{\label{fig2}
 Star-triangle transformation.  Left to right: star, triangle, duality 
 orientation.} \end{figure}

We illustrate explicitly our argument using the usual star-triangle transformation
shown in Fig.\ 2.  Here we will assume that the bond
occupation probabilities are $p_1$, $p_2$ and $p_3$ on the 
star (the original cell), and $P_1$, $P_2$, and $P_3$ on the
triangle (the dual cell).  Then we have explicitly for Eqs.\
(\ref{all}--\ref{none})
\begin{eqnarray}
p_1 p_2 p_3 &=& P_1 P_2 P_3 + P_1 P_2 Q_3 + P_1 P_3 Q_2 + P_2 P_3 Q_1 \label{allTri} \\
p_1 p_2 q_3 &=& Q_1 Q_2 P_3  \quad (\times 3) \label{oneTri} \\
q_1 q_2 q_3 &+& q_1 p_2 p_3 + q_2 p_1 p_3 + q_3 p_1 p_2 =  Q_1 Q_2 Q_3 \label{noneTri}
\end{eqnarray}
where $q_i = 1 - p_i$ and $Q_i = 1 - P_i$.
The two-point relation (\ref{oneTri}) is identically satisfied if $P_i = 1-p_i$.
 Then, there follows
from either (\ref{allTri})
or (\ref{noneTri}) the condition
$p_1 p_2 p_3 - p_1 p_2 - p_1 p_3 - p_2 p_3 + 1 = 0$,
which for $p_1 = p_2 = p_3 = p$ gives $p^3 - 3 p^2 + 1 = 0$
or $p_c = 1 - 2 \sin \pi/18 = 0.652704\ldots$ for bond percolation on the honeycomb
lattice \cite{SykesEssam}.

Above we repeated the complete argument to find $p_c$, working out probabilities
on the dual as well as the original system.  But using (\ref{ziff}),
we could just as well equate $P(A,B,C)=p_1 p_2 p_3$ to
$P(\overline A, \overline B, \overline C)
= q_1 q_2 q_3 + q_1 p_2 p_3 + q_2 p_1 p_3 + q_3 p_1 p_2$ to find the same
result without resorting explicitly to the dual lattice.

\begin{figure}
 \includegraphics[width=3in]{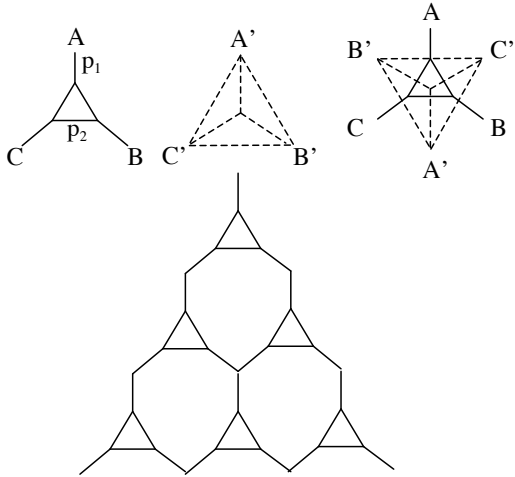} \caption{\label{fig3}
Cell, dual-cell, duality construction, and representation for 
Scullard's ``martini" lattice (bond percolation).} \end{figure}

\section{The martini lattice (bond percolation)}
Now we consider some new systems.  First we consider the unit cell shown
in Fig.\ 3, which is a triangle within a star.  As shown
in that figure, this cell forms precisely Scullard's
martini lattice, which is a honeycomb lattice with triangles inserted in every other
site.  Here we are considering bond percolation on that lattice.
We will assume that the outer three bonds are each occupied with
an equal probability $p_1$, and the inner three bonds 
are each occupied with probability $p_2$.
Then
a straightforward calculation shows that 
\begin{eqnarray}
P(A,B,C) &=& p_1^3(p_2^3 + 3 p_2^2 q_2)  \label{mbondall}\\
P(A,B,\overline C) &=& p_1^2 q_1 (p_2 + q_2 p_2^2)
 + p_1^3 (p_2 q_2^2) \label{mbondone}\\
 P(\overline A, \overline B, \overline C)
&=& q_1^3 + 3 q_1^2 p_1 + 3 q_1 p_1^2 (q_2^3 + 2 p_2 q_2^2)
+ p_1^3 (q_2^3) \label{mbondnone}
\end{eqnarray}
Equating
(\ref{mbondall}) and (\ref{mbondnone}), we find 
\begin{equation}
2 p_1^3 p_2^3 - 3p_1^3 p_2^2 - 3 p_1^2 p_2^3 + 3 p_1^2 p_2^2 
+ 3 p_1^2 p_2 - 1 = 0
\label{mbond}
\end{equation}
When $p_1 = 1$, this corresponds to a triangular lattice,
and when $p_2 = 1$, a honeycomb lattice.  When $p_1 = p_2 = p$,
we have 
$(2 p^2 - 1)(p^4 - 3 p^3 + 2 p^2 + 1 ) = 0$
which yields $p_c = 1/\sqrt{2}$, a new result.  Interestingly,
this is
the same value as the site percolation threshold
Scullard found for the ``A" lattice discussed below
(which is {\it not} the covering lattice
of the martini lattice).

\begin{figure}
\includegraphics[width=3in]{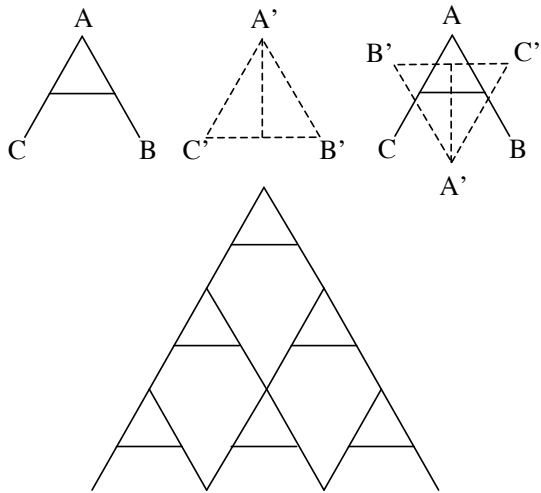} \caption{\label{fig4}
Cells and construction for bond percolation on the ``A" lattice.} \end{figure}

 It is instructive to 
 also work this out explicitly using the dual cell, which 
is a star within
 a triangle as shown in Fig.\ 3.  Here $P_1$ is the 
 occupancy of the outer bonds and $P_2$ is that of
 the inner bonds. We find
 \begin{eqnarray}
 P'(A,B,C) &=& P_1^3 + 3 P_1^2 Q_1 + 3 P_1 Q_1^2
    (P_2^3 + 2 P_2^2 Q_2) + Q_1^3 (P_2^3)\\
 P'(A,B,\overline C) 
    &=& P_1 Q_1^2 (Q_2 + P_2 Q_2^2) + Q_1^3 (P_2^2 Q_2) \\
 P'(\overline A,\overline B,\overline C) &=& 
    Q_1^3(Q_2^3 + 3 Q_2^2 P_2)
 \end{eqnarray}
 Then
 $ P(A,B,\overline C) = P'(A,B,\overline C)$
  if $Q_i = p_i$, and 
equating either  $P(A,B,C) = P'(A,B,C)$ or
$P(\overline A,\overline B,\overline C) = P'(\overline A,\overline B,\overline C)$ leads to the 
threshold given by (\ref{mbond}).

Now, we can see that the transformation shown in Fig.\ 3 of the
star-within-a-triangle to a triangle-within-a-star is analogous to the
star-triangle transformation, but intrinsically different in that 
that transformation of Fig.\ 3 cannot be accomplished by applying
the usual star-triangle transformation.  This is an example
of a more general cell/dual-cell transformation
that can be applied to percolation problems, where the cell can
be any graph connecting three vertices, and perhaps containing
correlated bonds.

\section{The ``A" and ``B" lattices (bond percolation)}
In the same way, we can find $p_c(\hbox{bond})$ for the other two lattices considered
by Scullard.  In Fig.\ 4 we
show what we call the ``A" lattice, since the basic cell has the shape
of an A.  It is equivalent to the cell of the martini lattice with
the upper bond removed or made occupied with probability 1. 
For simplicity, we assume all bonds are occupied with 
equal probablity $p$.
We find
\begin{eqnarray}
P(A,B,C) &=& p^2(p^3 + 3 p^2 q)  \\
P(\overline A, \overline B, \overline C)
&=& p^2 (q^3) + 2 p q (q ^ 3 + 2 p q^2) + q^2
 \end{eqnarray}
Equating
the two, we find
$
p^5 - 4 p^4 + 3 p^3 + 2 p^2 - 1  = 0
$
, whose solution gives $p_c(\hbox{bond})
 = 0.625457\ldots$.  This threshold is also new.

Fig.\ 4 also shows the dual cell --- a triangle with a vertical
line in it --- and once again, this defines a new cell/dual-cell
transformation, and one can verify directly that 
$ P(A,B,\overline C) =  P'(A,B,\overline C)$ with $P = 1-p$,
and rederive the above result for $p_c$.

\begin{figure}
 \includegraphics[width=3in]{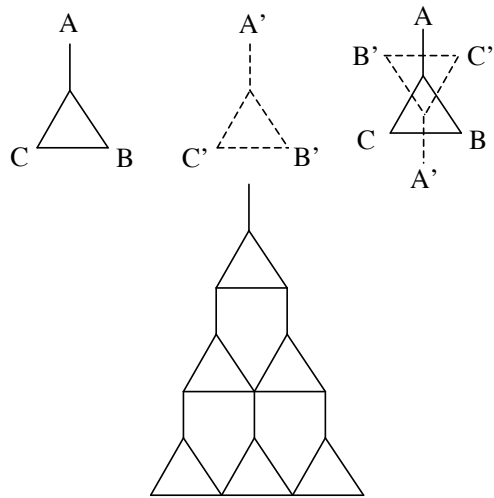} \caption{\label{fig5}
Construction for bond percolation on the ``B" lattice.} \end{figure}

Scullard's third lattice is generated by the cell shown in Fig.\ 5, 
which corresponds to the martini generator with two bonds
removed.  We call this the ``B" lattice, because the appearance
of B's when rotated by 90$^\circ$,
and also because it follows the ``A" lattice.
Because this lattice is self-dual, its $p_c(\hbox{bond})$ equals 1/2 as 
noted by Scullard.
This result is also borne out by (\ref{ziff}), since here
\begin{eqnarray}
P(A,B,C) &=& p (p^3+ 3 p^2 q) \\
P(\overline A, \overline B, \overline C)
&=& q (q^3 + 3 q^2 p)
 \end{eqnarray}
Equating
the two, one indeed finds 
$
(2 p - 1) (p^2 - p - 1) = 0
$, implying $p_c = 1/2$.

\begin{figure}
 \includegraphics[width=3in]{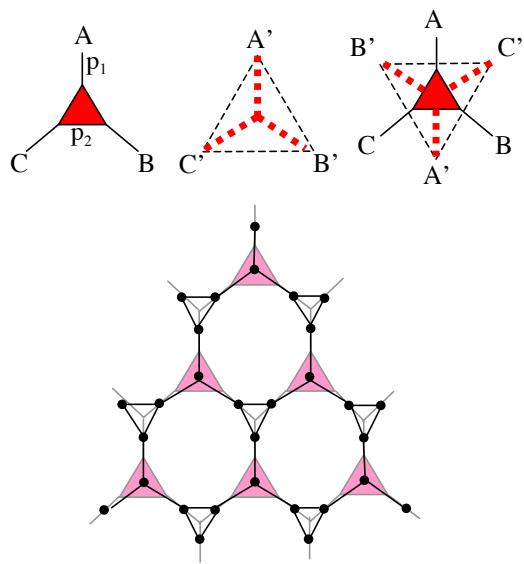} \caption{\label{fig6}
Construction for site percolation on the martini lattice, which
is the covering lattice shown in the lower part of the figure (color online).} \end{figure}

\section{Site percolation}

Now, we turn to some site percolation models. 
In general, some (but not all)
site percolation lattices can be created
by constructing the covering lattice of a bond problem.
Here we extend this procedure by also allowing for some of the bonds
to be correlated.  Thus in Fig.\ 6 we show a unit cell similar
to that of Fig.\ 3 except that now the three central bonds 
are all simultaneously occupied with probability $p$, 
and all vacant otherwise.  This is indicated by coloring the
triangle enclosed by those bonds.  Note that this triangle can also be thought
of as simply an independently occupied site, showing that this is essentially 
a site-bond problem, in which context
this problem was first solved by Kondor \cite{Kondor}.  Related
correlated site-bond problems were also
considered by Hu \cite{Hu} and Wu \cite{Wu}.
In a somewhat different context, Kunz and Wu 
also used the idea of creating site percolation by 
a group of correlated bonds in bond percolation system
\cite{KunzWu}.

Let $p_1$ be the occupancy of the three outer bonds,
and $p_2$ be the simultaneous occupancy of all three
central bonds. 
The covering lattice (which corresponds to the equivalent
pure site system)
is exactly the martini lattice, as shown
in Fig.\ 6.  We find
\begin{eqnarray}
P(A,B,C) &=& p_1^3 p_2 \\
P(\overline A, \overline B, \overline C)
&=& q_2 + p_2(q_1^3 + 3 p_1 q_1^2)
 \end{eqnarray}
 and setting these equal yields 
 $(p_1^3 - 3 p_1^2) p_2 + 1 = 0 $     
as given by Kondor \cite{Kondor}.
Setting $p_1 = p_2 = p$ yields
 $p^4 - 3p^3 + 1 = 0$ 
or $p_c = 0.764826\ldots$ for site percolation on the martini
lattice, as 
 found by Scullard.  Here, the dual lattice is a triangle with a correlated
 star in the center, and one can verify directly that the two-point functions
 on the lattice and dual lattice are equal when $P_i = 1 - p_i$.

\begin{figure} 
 \includegraphics[width=3in]{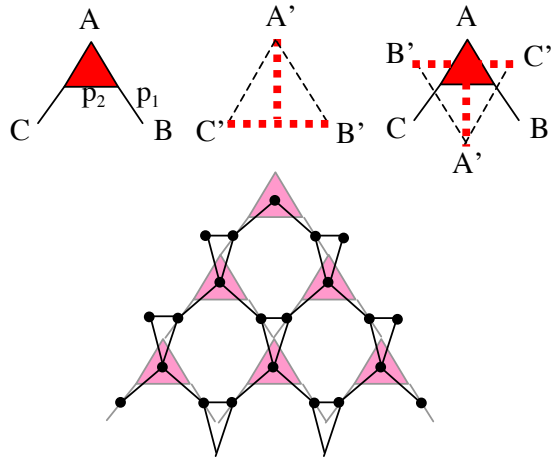} \caption{\label{fig7}
Construction for site percolation on the ``A" lattice (color online).}
\end{figure}

For site percolation on the ``A" lattice 
the cell/dual cell combination is shown in Fig.\ 7.
 Here we have 
\begin{eqnarray}
P(A,B,C) &=& p_1^2 p_2 \\
P(\overline A, \overline B, \overline C)
&=& q_2 + p_2 q_1^2
 \end{eqnarray}
 which yields
 $2 p_1 p_2 - 1 = 0$.  When $p_1=p_2=p$ this yields
 $p_c = 1/\sqrt{2}$ as found by Scullard.  For site percolation on the ``B" lattice, as shown in Fig.\ 8, we have 
$P(A,B,C) = p_1 p_2$ and
$P(\overline A, \overline B, \overline C) = q_2$
yielding $(p_1 + 1) p_2 - 1 = 0$.  For $p_1 = p_2 = p$,
this yields $p_c = (\sqrt{5}-1)/2 = 0.618034\ldots.$
as found by Scullard.

\begin{figure}
\includegraphics[width=3in]{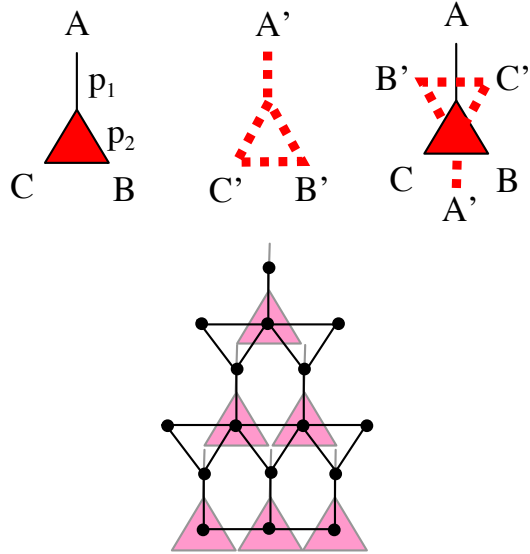} \caption{\label{fig8}
Construction for site percolation on the ``B" lattice (color online). }
\end{figure}
 
Likewise, one can create other cells that generate additional lattices where 
 $p_c$ can be found exactly.  By making the cell simply a shaded triangle
 (all three bonds occupied together), we create the site-triangular problem.
A cell of just two bonds creates the bond-square problem.  A cell with three 
correlated triangles touching creates the site-kagom\'e problem.  In this
way, all known thresholds can be easily calculated, and an infinite number 
of new ones can be created, with additional lattices
tending to be more and more intricate.  Unfortunately, the method does
not appear to work for some of the more notorious unsolved
systems: site percolation on the square and honeycomb lattices,
and bond percolation on the kagom\'e lattice.

An example of an intrinsically correlated system is one in which 
the cell is a simple triangle of bonds but where we require
that in each cell at least one bond is occupied, so that
$P(\overline A, \overline B, \overline C) = 0$.  Eq.\ (\ref{ziff})
implies that at criticality $P(A,B,C) = 0$ also, so that the
only non-zero correlation is $P(A,B,\overline C)$ and permutations,
meaning that the critical point corresponds to all cells having
exactly one occupied bond (in either a random or biased location), yielding
$p_c = 1/3$.  If only two of the three bonds can be occupied,
then the threshold corresponds to exactly one of those
two being occupied ($p_c = 1/2$), and we find 
a Scheidegger's river network model \cite{Scheidegger},
many of whose properties have been studied
and solved (e.g., \cite{TakayasuNishikawaTasaki,Huber}).
These are examples of fixed number or canonical
percolation models.

\begin{figure}
 \includegraphics[width=5in]{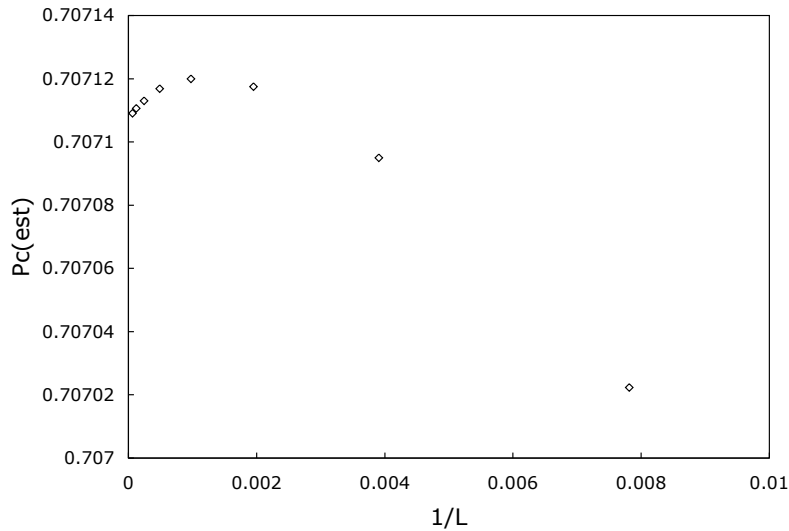} \caption{\label{plot}
Hull-gradient simulation results for site percolation on the ``A" 
lattice, confirming Scullard's prediction $p_c = 1/\sqrt{2} \approx 0.707107$.
Error bars are smaller than the symbol size.}
\end{figure}

\section{Numerical test for site percolation on the ``A" lattice"}
Finally, we report the results of a numerical test
of $p_c(\hbox{site})$ for the ``A" lattice, whose threshold was predicted
to be $1/\sqrt{2} = 0.70710678\ldots$ \cite{Scullard}.
We used
the hull-gradient method \cite{ZiffSapoval,SudingZiff},
with the lattice represented on a square lattice as 
$2\times 2$ squares with a diagonal in the left-lower
corner.  We considered systems of width
$L = 128, 256, \ldots 16384$ with the gradient $= 1/L$.  The estimates of $p_c$
(equal to the ratio of occupied to total sites in the hull)
are plotted in Fig.\ \ref{plot} vs.\ $1/L$, with $10^{12} - 10^{13}$
sites generated for each $L$. 
For $L = 16384$, we find $p_c(\hbox{est}) = 0.7071090$, and
the data extrapolate to the expected value as  $L \to \infty$ 
with a statistical error of about $\pm 0.000001$.
Note that there are
rather unusual finite-size effects for this system (perhaps
related to the orientation/representation of the lattice)
as
the estimates first increase and then decrease as $L$ increases.

We did not check any of the other predictions numerically:
because the cell/dual-cell argument is so compelling, 
there seems to be little doubt that these results
are correct.

\section{Conclusions and Discussion}
We have shown that the well-known star-triangle
transformation usually applied to honeycomb/triangular
lattices is a special case of a generalized transformation
where triangular cells in a system are individually
replaced by their duals. When the triangles themselves
form a self-dual system as in Fig.\ 1a, then the generalized criterion
for percolation (\ref{ziff}) follows.  Interestingly,
to apply this criterion one does not need to 
explicitly apply the star-triangle or duality
transformation, as the criterion involves 
connectivity probabilities on just the original system.

Note that Fig.\ 1a is but one arrangement of 
the triangular cells that is self dual.
Other arrangements can also be devised, and
these will be pursued in future work.

The basic triangular cells can possess correlations
in their connectivities, and we have shown that 
correlations can be the result of having an uncorrelated
bond system that is more complicated than a simple
triangle or star.  Site percolation can also be included by
adding triangles of correlated bonds.  In this
way, we have provided a direct proof of Scullard's
new results for the site percolation threshold for
the martini lattice, and its two contractions,
the ``A" and ``B" lattices.  We derived
the bond thresholds of these lattices also, which
represents two additional new thresholds.   Note that
these three lattices are the only known cases, besides
the triangular lattice, where both site and bond
percolation thresholds can be found exactly.  
Other soluble lattices can easily be generated by
putting additional bonds on the basic cells, although
the lattices thus formed take on more and more the look
of ``decorated" rather than regular lattices.

It turns out that the criterion that we derived,
Eq.\ (\ref{ziff}), is closely related to work 
done many years ago on the Potts model in
correlated systems \cite{fywuPrivateComm}. 
In fact, (\ref{ziff}) is a special case of a criterion
for being at the critical point of the correlated Potts
model given by Wu and co-workers \cite{WuLin,WuZia,MaillardRolletWu,KingWu}.
However, evidently the percolation limit of that
work was never explored.  

Finally, after this paper was submitted for publication,
we came across a recent preprint of work by
Chayes and Lei \cite{Chayes}
in which the generalized criterion
 (\ref{ziff}) was also derived.  The focus of
their work is quite different,
and these authors do not use that
criterion to find the thresholds
of new lattices as we have done here.

\section{Acknowledgments}
  The author thanks Chris Scullard
 for sending an advance copy of his paper and for
 a stimulating exchange of email correspondence,
 Greg Huber for discussions on river networks,
 and F. Y. Wu for comments and 
 for assistance on the relation
 to previous work on the correlated Potts model.  The author
 acknowledges financial
 support from NSF grant DMS-0244419.  


\begin{thebibliography}{99}

\bibitem{StaufferAharony} D. Stauffer and A. Aharony, {\sl Introduction to Percolation Theory}
(Taylor and Francis, 1994).

\bibitem{SykesEssam} M. F. Sykes and J. W. Essam, J. Math. Phys. {\bf 5}, 1117 (1964).

\bibitem{Wierman} J. C. Wierman, J. Phys. A {\bf 17}, 1525 (1984).

\bibitem{Scullard} C. Scullard, 
``Exact site percolation thresholds using a site-to-bond transformation and the star-triangle transformation,"  Phys. Rev. E (to appear) (2006)

\bibitem{Kondor} I. Kondor, J. Phys. C {\bf 13}, L531 (1980).

\bibitem{Hu} C. K. Hu, Phys. Rev. B {\bf 29}, 5103 (1984).

\bibitem{Wu} F. Y. Wu, J. Phys. A {\bf 14}, L39 (1981).

\bibitem{Scheidegger} A. E. Scheidegger, Bull. Int. Assoc. 
Sci. Hydrology {\bf 12}, 15 (1967).

\bibitem{TakayasuNishikawaTasaki} H. Takayasu, I. Nishikawa,
and H. Tasaki, Phys. Rev. A {\bf 37}, 3110 (1988).

\bibitem{Huber} G. Huber, Physica A {\bf 170}, 463 (1991).

\bibitem{WuLin} F. Y. Wu and K. Y. Lin, J. Phys. A {\bf 13}, 629 (1980).

\bibitem{WuZia} F. Y. Wu and R. K. P. Zia, J. Phys. A {\bf 14}, 721 (1981).

\bibitem{KingWu} C. King and F. Y. Wu, Int. J. Mod. Phys. B {\bf 11},
51 (1997).

\bibitem{MaillardRolletWu} J. M. Maillard, G. Rollet, and F. Y. Wu,
J. Phys. A {\bf 26}, L495 (1993).

\bibitem{KunzWu} H. Kunz and F. Y. Wu, J. Phys. C {\bf 11}, L1, L357 (1977).

\bibitem{ZiffSapoval} R. M. Ziff and B. Sapoval,
J. Phys. A {\bf 19} L1169 (1986).Ê

\bibitem{SudingZiff} P. N. Suding and R. M. Ziff, Phys. Rev. E {\bf 60}, 275 (1999).

\bibitem{fywuPrivateComm} F. Y. Wu, private communication

\bibitem{Chayes} L. Chayes and H. K. Lei, ``Random cluster models on the
triangular lattice," preprint arxiv:cond-mat/0508253, 10 August 2005.

\end{thebibliography}
\end{document}